\documentclass[onecolumn]{revtex4}
\usepackage{color}
\usepackage{graphicx}
\usepackage{amsmath}
\usepackage{float}
\usepackage[english]{babel}

\begin{document}

\title{Existence of Korteweg-de Vries Solitons and Relevance of Relativistic\\ Effects in a Dusty Electron-Ion Plasma}

\author{Maricarmen A. Winkler}
\email{maricarmen.winkler@uchile.cl (corresponding author)}
\affiliation{Departamento de F\'isica, Facultad de Ciencias, Universidad de Chile, Casilla 653, Santiago, Chile}

\author{V\'{\i}ctor Mu\~noz}
	\email{vmunoz@macul.ciencias.uchile.cl}
\affiliation{Departamento de F\'isica, Facultad de Ciencias, Universidad de Chile, Casilla 653, Santiago, Chile}

\author{Felipe A. Asenjo}
\email{felipe.asenjo@uai.cl}
\affiliation{Facultad de Ingenier\'ia y Ciencias,
Universidad Adolfo Ib\'a\~nez, Santiago 7491169, Chile.}

\date{\today}
\begin{abstract}
Nonlinear effects in the propagation of perturbations in a dusty electron-ion plasma is studied, considering fully relativistic wave motion. A multifluid model is considered for the particles, from which a KdV equation can be derived. In general, two different soliton solutions are found depending on the kind of dispersion of the KdV equation. We study when the dispersion coefficient of this equation is positive. In this case, two kind of behavior are possible, one associated with a slow wave mode, another with a fast wave mode. It is shown that, depending on the value of the system parameters, compressive and/or rarefactive solitons, or no soliton at all, can be found and that relativistic effects for ions are much more relevant than for electrons. It is also found that relativistic effects can strongly decrease the soliton amplitude for the slow mode, whereas for the fast mode they can lead to compressive-rarefactive soliton transitions and vice versa, depending on the dust charge density in both modes.
\end{abstract}

\pacs{}

\maketitle

\section{Introduction}\label{introduction}

Dusty plasmas have been the subject of a large amount of research during recent decades, due to their relevance to model laboratory, space and astrophysical systems \cite{Shukla_s,Verheest_e}. They have been found in the vicinity of various objects in the solar system such as the Moon \cite{Vaverka}, comets \cite{Horanyi_a,Cremonese,Chen,Naeem}, planets \cite{Havnes,Khalisi,Kruger,Horanyi} and in the interplanetary medium \cite{Mann_a}. Direct observations by the Ulysses spacecraft confirmed the existence of interstellar dust grains in the solar system, with a time-dependent flow near the Sun~\cite{Altobelli,Mann}. Also, several works have discussed the relevance of dust particles for the plasma dynamics in astrophysical objects~\cite{Lee_f,Draine,Mahmoodi,Fry}. 
Dust particles are also relevant for the study of fusion plasmas, leading to an increasing interest to understand how dust interacts with the plasma. These grains may negatively impact the performance of a tokamak device, although there is research on their possible positive effects~\cite{Bacharis,Rudakov} as well. Various experiments in magnetized dusty plasmas~\cite{Thomas,Fortov} have been carried out, in order to study them under controlled conditions. 

Various works have been devoted to the study of soliton formation and evolution in dusty plasmas~\cite{Verheest_f,Ma}, including experimental~\cite{Nakamura_a,Bandyopadhyay} and analytical treatments under different regimes, such as the presence of a background magnetic field~\cite{Malik}, generalized particle distributions~\cite{Baluku,ElBedwehy} or presence of negatively and positively charged dust particles~\cite{Sayed,Mannan}. If particle velocities are large, relativistic effects may modify the properties of solitons. For instance, in the work of El-Shamy {\it et al.\/}~\cite{ElShamy} is shown that, in dust ion-acoustic soliton collisions, higher-order phase shifts decrease due to weakly relativistic effects. On the other hand, in the work of Zobaer {\it et al.\/}~\cite{Zobaer}, the propagation of dust ion-acoustic waves is studied in a degenerate plasma, finding that soliton amplitude and width increases when the electrons are ultrarelativistic degenerate, as compared with the case of nonrelativistic degenerate electrons, when ions are considered nonrelativistic degenerate.
Kalita and Choudhury~\cite{Kalita} studied the existence and properties of solitons a  in a weakly relativistic, unmagnetized electron-ion-dust plasma. Soliton amplitude and profiles are shown as a function of the species velocities and the electron temperature. It is shown that the species velocities determines the existence of both compressive and rarefactive relativistic solitons, or the existence of only rarefactive solitons, depending on the precise values of the other parameters. However, this work is limited to the weakly relativistic regime, and relativistic velocities may be needed for the existence of solitons of significant amplitude in other regions of the parameter space. 

The fully relativistic regime is also worth to be considered, as interesting new effects may appear when velocities are arbitrarily large. For instance, relativistic velocities may increase the plasma transparency to electromagnetic waves, and introduce cutoffs in the Alfv\'en mode in electron-positron plasmas, so that Alfv\'en waves can only propagate for a finite band of wavenumber~\cite{Asenjo_a}. This, in turn, has consequences on the existence of solitons associated to each wave mode  (see for example Refs.~\cite{Asenjo_d,Lopez_d}). All these works focused on the effect of arbitrarily large particle velocities and temperatures, without considering the presence of dust. 

Given the discussion above, it would be interesting to extend previous works to study the existence and properties of solitons in a plasma where both dust and relativistic velocities in all species should be considered. This could be relevant in various systems involving laboratory, space, and especially astrophysical plasmas, where extreme conditions are expected. For instance, star collapse or supernova explosions may lead the particles to relativistic energies, where plasma jets are injected into the dusty interstellar medium~\cite{Taherimoghadam,Higdon,Saini}.
Also, the collision of clouds in the dusty torus around accretion disks near black holes~\cite{Mehlhaff,Nenkova,Muller,Singh,Kynoch}, may create a population of particles accelerated to relativistic velocities~\cite{Wang_f}.

Thus, in this work, we study solitons in the fully relativistic regime, beyond the weakly relativistic approximation shown in Ref.~\cite{Kalita}. We find that fluid equations also lead to a Korteweg-de Vries (KdV) equation for the electrostatic potential, and that solitons can propagate in the system. We
consider fully relativistic effects in the wave field consistently and study the parameter space to find out under what conditions KdV solitons are possible, their width and amplitude, and the relevance of relativistic effects in electrons and ions. The paper is structured as follows. In Sec.~\ref{model} the model equations and basic assumptions are presented. Then, in Sec.~\ref{KdV} a KdV equation is obtained from the model. In Sec.~\ref{numerical} the equations are solved numerically, so that the existence of solitons and their characterization can be studied as a function of various plasma parameters, in particular the ones related to the dust. Given that our results extend previous results dealing with weakly relativistic effects, in Sec.~\ref{relativistic} we focus on how relativistic effects modify the soliton properties, compared with the nonrelativistic results. This is done by comparing the nonrelativistic, weakly relativistic and fully relativistic cases. Finally, in Sec.~\ref{summary} results are summarized and discussed.


\section{Model equations}\label{model}

To study the dynamics and soliton propagation of an unmagnetized, homogeneous electron-ion-dust plasma, we will use fluid theory considering negatively charged, nonrelativistic dust particles and relativistic electrons and ions.
Then, the system is described by the continuity equation
\begin{align}
\frac{\partial}{\partial t'}\left(n'_j \gamma_j\right) &+ \frac{\partial}{\partial z'}\left(n'_j \gamma_j v'_j\right) = 0\ , \label{eq:1} 
\end{align}
where $j$ denotes the species index: $j=i$ for ions, $j=e$ for electrons and $j=d$ for dust particles. Besides, $v'_j,$ $n'_j$,  and $\gamma_j$ are normalized quantities for the velocity, rest-frame number density of each fluid and relativistic Lorentz factor
\begin{equation}
\gamma_j=\left(1-\frac{{v'_j}^2}{{c'}^2}\right)^{-1/2} = \left(1-\frac{{v_j}^2}{c^2}\right)^{-1/2}  \ ,
\end{equation}
respectively. 
Here, the number densities are normalized by the equilibrium ion density $n_{i0}$,  distances by $\lambda_D=\left(k_BT_e/4\pi n_{i0}e^2\right)^{1/2}$, velocties by $v_D=\left(k_BT_e/m_d\right)^{1/2}$, and time by $t'=\left(4\pi e^2 n_{i0}/m_d\right)^{-1/2}$, where $e$ is the magnitude of the electron charge, $k_B$ is the Boltzmann constant, $T_e$ is the electron plasma temperature, and $m_d$ is the mass of the dust particles.
Besides, we have 
the momentum equation 
\begin{align}
n'_j\gamma_j Q_j\left(\frac{\partial}{\partial t'} + v'_j \frac{\partial}{\partial z'}\right)\left(\gamma_j v'_j\right) &=\eta_j n'_j\gamma_j\frac{\partial\phi'}{\partial z'}-\alpha_j\frac{\partial n'_j}{\partial z'} \ , \label{eq:2}
\end{align}
where $\phi'$ is the    electric potential  normalized by $k_BT_e/e$. 
In addition, the
following quantities have been defined: the negative of the atomic number $\eta_j$, a normalized temperature $\alpha_j=T_j/T_e$, a normalized mass $Q_j=m_j/m_d$ and as dust particles are nonrelativistic, $\gamma_d=1$. Notice that, for the system considered in this paper, $\eta_e= Z_e =1$, $\eta_i =-Z_i = -1$, $\eta_d=Z_d$, $\alpha_d = 0$, $\alpha_e=1$, and $Q_d=1$. Also, a value of $\alpha_i=0.1$ was chosen for ions, as suggested by Kalita and Das~\cite{Kalita2014}.
Finally, we complete the system with the Poisson equation  
\begin{align}
\frac{\partial^2\phi'}{\partial {z'}^2} &= \ \sum_j \eta_j n'_j \gamma_j\ . \label{eq:3}
\end{align}

These fluid equations can be derived from the fully relativistic fluid equation written in covariant form (see, for instance, Asenjo {\it et al.\/}~\cite{Asenjo_a}). Since the energy-momentum tensor has to be written in terms of the enthalpy in the rest frame~\cite{Misner,Landau}, we have taken care to write our equations in terms of the rest-frame density, although they can easily be changed to the laboratory frame density $n_L'=\gamma_j n_j'$.
It is worth noticing that the equivalent equations in Kalita and Choudhury~\cite{Kalita} are also written in terms of the rest-frame densities (as seen in the $\gamma_j n_j$ factors in the continuity equation, for instance), but, if so, there is an extra $\gamma$ factor in the pressure term of Eq.~(4) in Ref.~\cite{Kalita}.

In order to close the system of Eqs. \eqref{eq:1}--\eqref{eq:3}, the equation of state $p_j= n_j k_B T_j$ for an ideal gas will be used. 


\section{K\lowercase{d}V solitons}\label{KdV}

For small and finite nonlinear perturbations, we use the reductive perturbation method, introducing stretched coordinates $\xi=\varepsilon^{1/2}\left(z'-Mt'\right)$ and $\tau =\varepsilon^{3/2}t'$, where $M$ is the phase velocity of the wave and $\varepsilon \ll 1$ is a parameter that describes the nonlinear response of the system. According to this, we expand for all species $n'_j =  n'_{j0} + \varepsilon n'_{j1} + \varepsilon^2 n'_{j2} + \cdots$, $
v'_j  =  v'_{j0} + \varepsilon v'_{j1} + \varepsilon^2 v'_{j2} + \cdots $, and $\phi'  =  \varepsilon \phi'_{1} + \varepsilon^2 \phi'_{2} + \cdots $.
With these expansions, along with Eqs. \eqref{eq:1}--\eqref{eq:3}, we gather terms up to second order in $\varepsilon$. The zeroth order terms allow us to compute an expression for the electron density
\begin{equation}
n'_{e0}=\frac{\gamma_{i0}-Z_d n'_{d0}}{\gamma_{e0}} \ .
\end{equation}
Then,  the first order terms in the expansions of the continuity equation becomes
\begin{align}
\left(v'_{j0}-M\right)\frac{\partial n'_{j1}}{\partial\xi} + n'_{j0}\gamma_{j0}^2\bigg(1 -&\left.\frac{M v'_{j0}}{{c'}^{2}}\right)\frac{\partial v'_{j1}}{\partial\xi} = 0\ ,
\label{eq:5} 
\end{align}
where $\gamma_{j0}=(1-{v'_{j0}}^2/{c'}^2)^{-1/2}$ was used.
Similarly, 
the momentum equation at first order is
\begin{align}
n'_{j0}\gamma_{j0}^4\left(v'_{j0}-M\right)\frac{\partial
  v'_{j1}}{\partial\xi} - \frac{\eta_j}{Q_j}
n'_{j0}&\gamma_{j0}\frac{\partial \phi'_1}{\partial\xi}+ \frac{\alpha_j}{Q_j}\frac{\partial n'_{j1}}{\partial\xi}=0\ , \label{eq:6}
\end{align}
while the Poisson equation is
\begin{align}
\sum_j \gamma_{j0}\eta_j \bigg(n'_{j1}+ n'_{j0}&\gamma_{j0}^2\frac{v'_{j0}}{{c'}^2}v'_{j1}\bigg)=0 \ . \label{eq:7}
\end{align}
 Thereby, from the above equations we obtain expressions for $n'_{j1}$, $v'_{j1}$ and $M$, as
\begin{align}
n'_{j1} &= -\frac{n'_{j0}\gamma_{j0}\eta_j}{R_j}\left(1-\frac{M v'_{j0}}{{c'}^2}\right)\phi'_1\ ,\label{eq:8} \\
v'_{j1} &= \frac{\eta_j\left(v'_{j0}-M\right)}{\gamma_{j0}R_j}\phi'_1\ ,\label{eq:9}\\
0 &= \sum_j \frac{n'_{j0}\eta_j^2}{R_j}\ , \label{eq:10}
\end{align}
where $R_j$ is defined as
\begin{equation*}
R_j=Q_j \gamma^2_{j0} \left(v'_{j0}-M\right)^2 - \alpha_j\left(1-\frac{Mv'_{j0}}{c^2}\right)\ .
\end{equation*}

Next, we obtain the second order expressions for the continuity equation
\begin{align}
&\left(v'_{j0}-M\right)\frac{\partial n'_{j2}}{\partial \xi} +n'_{j0}\gamma^2_{j0}\bigg(1- \frac{Mv'_{j0}}{{c'}^2}\bigg)\frac{\partial v'_{j2}}{\partial \xi}+\gamma^2_{j0} \left(1 -\frac{Mv'_{j0}}{{c'}^2}\right)\frac{\partial}{\partial\xi}\left(n'_{j1}v'_{j1}\right) + \frac{\partial n'_{j1}}{\partial \tau}  \nonumber \\
&\phantom{.}+n'_{j0}\gamma^2_{j0}\left[\frac{2v'_{j0}}{{c'}^2} +\frac{\left(v'_{j0}-M\right)}{{c'}^2}\left(1 + 3\gamma^2_{j0}\frac{{v'_{j0}}^2}{{c'}^2}\right)\right]v'_{j1}\frac{\partial v'_{j1}}{\partial \xi} + n'_{j0}\gamma^2_{j0}\frac{v'_{j0}}{{c'}^2}\frac{\partial v'_{j1}}{\partial \tau}=0\ , \label{eq:11}
\end{align}
the momentum equation
\begin{align}
&n'_{j0}\gamma^4_{j0}\left(v'_{j0}-M\right)\frac{\partial v'_{j2}}{\partial \xi}+\frac{\alpha_j}{Q_j}\frac{\partial n'_{j2}}{\partial\xi}-\frac{n'_{j0}\eta_j}{Q_j}\gamma_{j0}\frac{\partial\phi'_2}{\partial\xi}+ n'_{j0}\gamma^4_{j0}\frac{\partial v'_{j1}}{\partial\tau}+ \gamma^4_{j0}\left(v'_{j0} - M\right)n'_{j1}\frac{\partial v'_{j1}}{\partial \xi}\phantom{.} \nonumber\\
&+n'_{j0}\gamma^4_{j0}\left[1 + 4\gamma^2_{j0}\frac{v'_{j0}}{{c'}^2}\left(v'_{j0}-M\right)\right]v'_{j1}\frac{\partial v'_{j1}}{\partial \xi}-\frac{\eta_j}{Q_j}\gamma_{j0}\left[n'_{j1}+n'_{j0}\gamma^2_{j0}\frac{v'_{j0}}{{c'}^2}v'_{j1}\right]\frac{\partial \phi'_{1}}{\partial\xi}  =0\ ,\label{eq:12}
\end{align}
and the Poisson equation
\begin{align}
\frac{\partial^2\phi'_1}{\partial \xi^2} &= \sum_j \eta_j\gamma_{j0}\left[n'_{j2} + n'_{j0}\gamma^2_{j0}
\frac{v'_{j0} v'_{j2}}{{c'}^2} +\gamma^2_{j0}\frac{v'_{j0}}{{c'}^2}\left(n'_{j1} v'_{j1}\right)
+n'_{j0}\gamma^2_{j0}\left(1+ 
 3\gamma^2_{j0}\frac{{v'_{j0}}^2}{{c'}^2}\right)\frac{{v'_{j1}}^2}{2{c'}^2}\right]\ . \label{eq:13} 
\end{align} 

These calculations represent an improvement and extension on the results presented by Kalita and Choudhury~\cite{Kalita}, as relativistic effects on the velocities have been kept consistently, so that we are able to consider relativistic zeroth-order velocities ($v'_{j0}$). 

Using Eqs. \eqref{eq:8}--\eqref{eq:10} along with \eqref{eq:11} and \eqref{eq:12}, we find expressions for $n'_{j2}$ and $v'_{j2}$. Using them, and the first order quantities $n'_{j1}$ and $v'_{j1}$, after algebraic manipulation of Eq.~\eqref{eq:13}, we are able to find the KdV equation for $\phi'_1$. This results to be 
\begin{equation}
\frac{\partial\phi'_1}{\partial\tau}+p\phi'_1\frac{\partial\phi'_1}{\partial\xi}+q\frac{\partial^3\phi'_1}{\partial\xi^3} = 0\ , \label{eq:14}
\end{equation} 
where $p=B/A$ is the nonlinear coefficient and $q=1/A$ the dispersion coefficient, considering that
\begin{align}
A &= \sum_j \frac{n'_{j0}\eta_j^2}{R^2_j}\left[\frac{v'_{j0}}{{c'}^2}\alpha_j - 2 Q_j\left(v'_{j0}-M\right)\right]\ , \label{eq:15} \\
B &= \sum_j \frac{n'_{j0}\eta_j^3 }{\gamma_{j0}R_j^3} \left\lbrace\rule{-0.1cm}{0.65cm}\right. \alpha_j \left[\rule{-0.1cm}{0.63cm}\right.  1-\frac{M v'_{j0}}{{c'}^2} + \frac{\gamma^2_{j0}}{{c'}^2}\left(v'_{j0}-M\right)^2\left.\rule{-0.1cm}{0.65cm}\right]- 3Q_j\gamma^4_{j0}\left(v'_{j0}-M\right)^2\left(1-\frac{Mv'_{j0}}{{c'}^2}\right)\left.\rule{-0.1cm}{0.65cm}\right\rbrace \ . \label{eq:16}
\end{align}

The KdV equation \eqref{eq:14} has two different (related) soliton solutions depending on the sign of the parameter $q$.

\subsection{Solitons for $q>0$}

Eq.~\eqref{eq:14} has a known soliton solution 
\begin{equation}
\phi'_1 = \phi_0\ \text{sech}^2\left(\frac{\eta}{\Delta}\right)\ , \label{eq:17}
\end{equation}
with $\eta=\xi -V\tau$, $\phi_0=3V/p$ and  $\Delta=\sqrt{4|q|/V}$, considering $V$ as the soliton speed.
This solution is only possible when the dispersion coefficient $q$ is positive.

\subsection{Solitons for $q<0$}

On the contrary, for this case, the soliton solution is
\begin{equation}
\phi'_1 = \frac{\phi_0}{2}\ \text{tanh}^2\left(\frac{\eta}{\sqrt{2}\Delta}\right)\ . \label{eq:18b}
\end{equation}
Then, when $q<0$, the amplitude of the soliton is reduced to half, and its width is increased by $\sqrt{2}$. However, both soliton solutions propagate with the same speed.

This solution has the opposite behavior to soliton \eqref{eq:17}. When soliton 
\eqref{eq:17} is compressive (rarefactive), soliton \eqref{eq:18b} is rarefactive (compressive).

\section{Numerical results}\label{numerical}

In order to obtain numerical solutions, we set values for the free parameters, namely normalized density $n_{j0}'$, velocity $v_{j0}'$, temperature $\alpha_j$, ion (electron) to dust mass ratio $Q_i$ $(Q_e)$, and soliton speed $V$. With this, Eq.~\eqref{eq:10} can be numerically solved for $M$. This allows to evaluate $p$ and $q$, and thus to obtain the amplitude, width and shape of the soliton. 

In general, Eq.~\eqref{eq:10} yields four solutions for $M$. However, depending on the free parameters, it can happen that all values are real, only two of them, or none. Thus, we first select all real solutions for $M$. Then, for each chosen value of $M$, we evaluate the dispersion coefficient $q$. 

In the following, we only focus on the case when $q>0$, and therefore, in the soliton solution \eqref{eq:17}. The same analysis can be straightforwardly repeated for soliton solution \eqref{eq:18b}.

Using the chosen values of $M$, we test if the resulting value of $q$ is real and positive, which allows us to select the physical values of $M$. Finally, we evaluate the corresponding values of the nonlinear coefficient $p$. It turns out that there are at most two physically relevant values of $M$, corresponding to two phase velocities. We can thus identify a slow and a fast mode.

\begin{figure}[h]\centering
\begin{tabular}{cc}
\includegraphics[width=6.15cm]{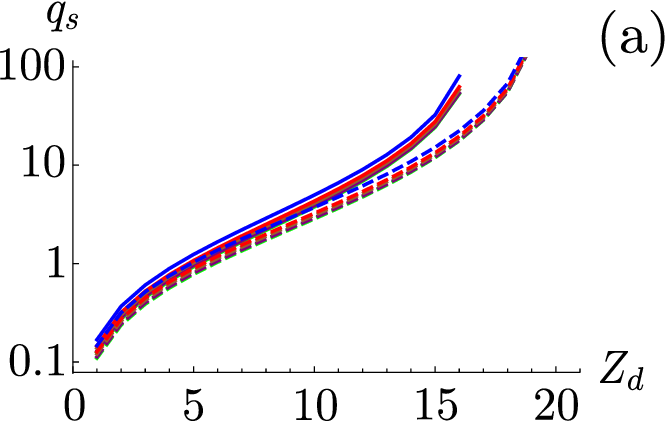} & \includegraphics[width=6.15cm]{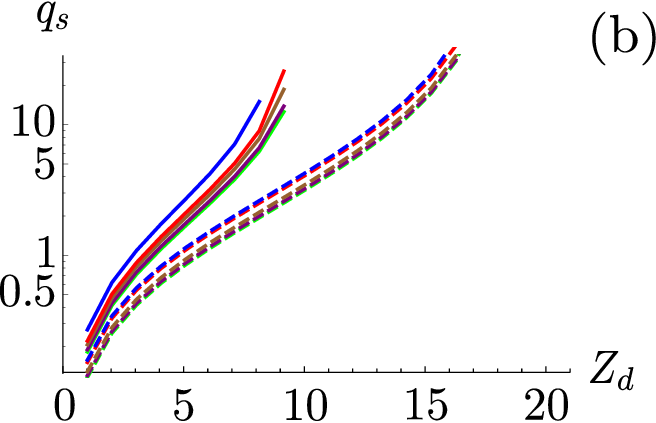} \\
\includegraphics[width=6.15cm]{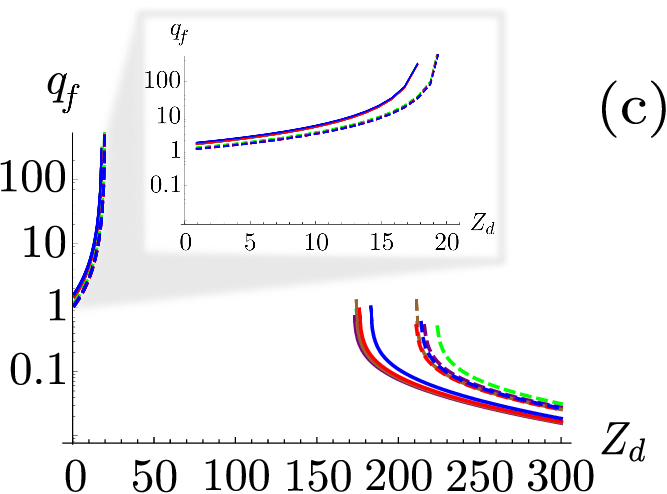} & \includegraphics[width=6.15cm]{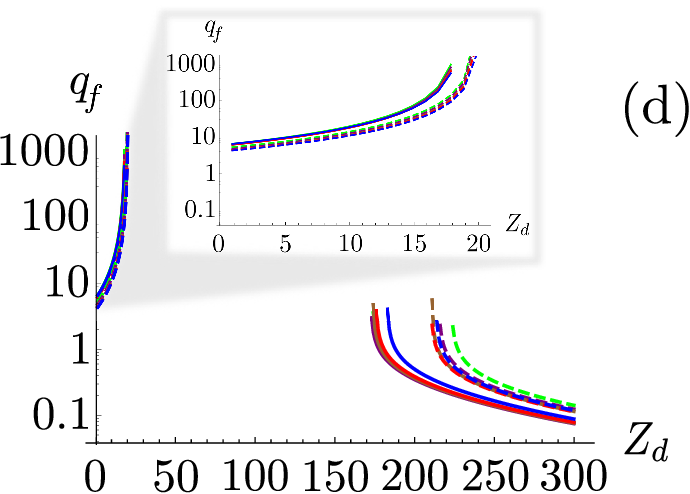}
\end{tabular}
\caption{Dispersion coefficient for the slow mode $q_s$ and fast mode $q_f$ versus dust charge $Z_d$, when $n'_{d0}=0.055$ and $c'=140$. For (a) and (c) the ion to dust mass ratio is $Q_i=0.1$ and for (b) and (d) is $Q_i=0.005$. Solid (dashed) lines correspond to $v'_{i0}=30$ ($v'_{i0}=60$), which represent weakly relativistic (relativistic) ions. Colors correspond to different electron velocities: $v'_{e0}=10$ (green), $v'_{e0}=30$ (purple), $v'_{e0}=50$ (brown), $v'_{e0}=60$ (red), and $v'_{e0}=80$ (blue).}
\label{fig1}
\end{figure}
 
Fig.~\ref{fig1} shows the $q$ coefficient when it is positive, for two values of ion to dust mass ratio, for several electron/ion velocities, and for each mode: $q_s$ for the slow mode, panels (a) and (b), and $q_f$ for the fast mode, panels (c) and (d). In particular, two ion velocities are considered to see the relevance of weakly relativistic ($v'_{i0}=30$) and relativistic ($v'_{i0}=60$) effects on $q$. For each $v'_{i0}$, several values of the electron velocities are considered, ranging from weakly relativistic to relativistic values ($v'_{e0}=10,\dots , 80$). 
We notice that the slow mode has a single branch, and that there is a maximum value of the dust charge for soliton existence that depends on ion velocity, as it can be seen in panels \ref{fig1}(a) and \ref{fig1}(b).

\begin{figure}[h]\centering
\begin{tabular}{cc}
\includegraphics[width=6.15cm]{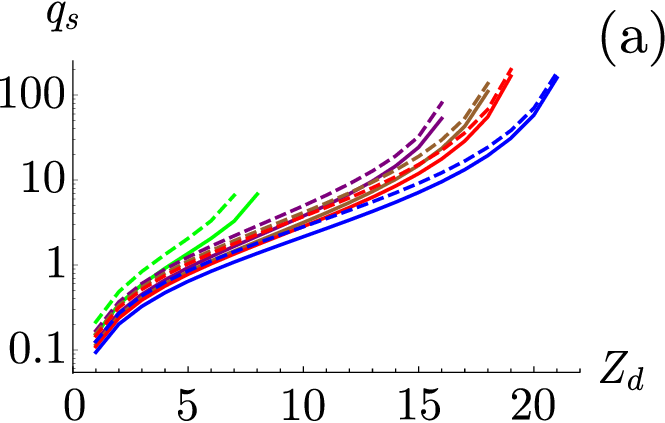} & \includegraphics[width=6.15cm]{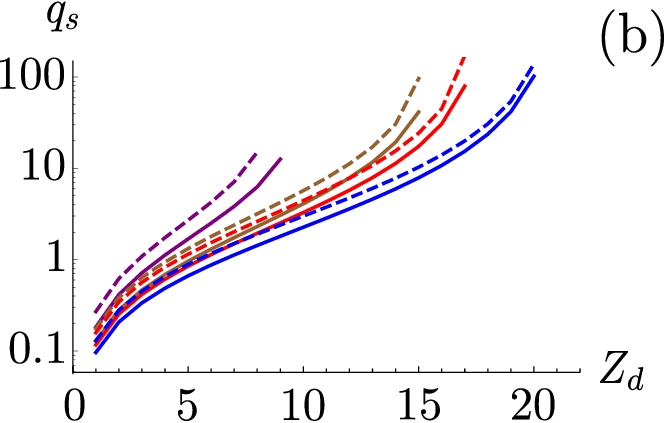} \\
\includegraphics[width=6.15cm]{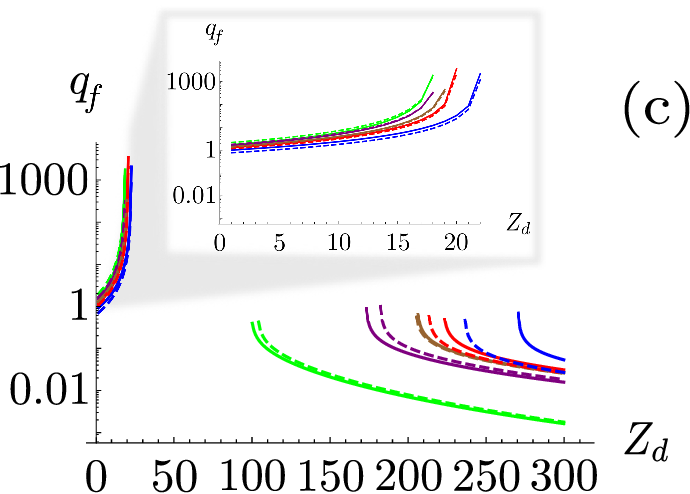} & \includegraphics[width=6.15cm]{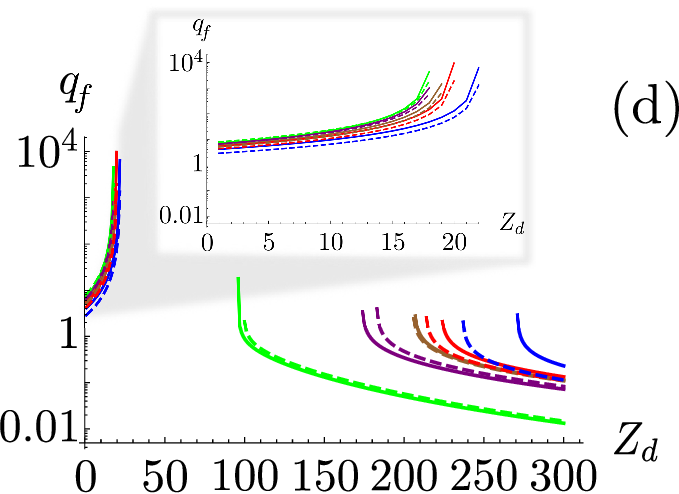}
\end{tabular}
\caption{Same as Fig. \ref{fig1}, but exchanging the choice of velocities for ions for those of electrons. In this case, solid (dashed) lines represent weakly relativistic (relativistic) electrons, $v'_{e0}=10$ ($v'_{e0}=80$). Colors correspond to different ion initial velocities: $v'_{i0}=10$ (green), $v'_{i0}=30$ (purple), $v'_{i0}=50$ (brown), $v'_{i0}=60$ (red) and $v'_{i0}=80$ (blue).}
\label{fig2}
\end{figure}

On the other hand, the fast mode has two branches, one for small values of $Z_d$ and another for large values of $Z_d$. The limits of the dust charge for both branches also depend on ion velocity and are separated by an intermediate region, where $q_f$ has complex values, shown in panels \ref{fig1}(c) and \ref{fig1}(d). Considering relativistic ion velocities, when $Z_d$ is small, the range of dust charge where solitons exist increases and the value of $q$ decreases, for both the slow and fast modes. When $Z_d$ is large in the fast mode, the value of $q$ increases, and the range of dust charge where solitons exist decreases. This is shown in the difference between solid and dashed lines in Fig.~\ref{fig1}. However, relativistic electron velocities (as seen in the different line color in Fig.~\ref{fig1}), do not have a major effect. Finally, we find that, for the studied parameter region, $Q_i$ clearly changes the region in $Z_d$ where the solitons exist for the slow mode [Figs.~\ref{fig1}(a),~(b)], but there is no appreciable effect for the fast mode [Figs.~\ref{fig1}(c),~(d)]. For both cases, though, the ion to dust mass ratio modifies the width of the solitons (notice the logarithmic scale in the $q$-axis). 

Fig.~\ref{fig2} is similar to Fig.~\ref{fig1}, except that two velocities for electrons are considered: one relativistic ($v'_{e0}=80$) and one weakly relativistic ($v'_{e0}=10$). For each $v'_{e0}$, several values of the ion velocity are considered, ranging from weakly relativistic to relativistic velocities ($v'_{i0}=10,\dots , 80$). Fig.~\ref{fig2} shows the same behaviour as Fig.~\ref{fig1}, the slow mode exists up to a maximum value of $Z_d$, whereas the fast mode has a forbidden region for intermediate values of $Z_d$. Relativistic effects on the electrons do not play a major role, as illustrated by the fact that dashed and solid lines of a given color are always similar. Instead, changing the ion velocity has the main effect on the mode: lines are separated by color in Fig.~\ref{fig2}. Changing the value of $Q_i$ has a less important effect than on Fig.~\ref{fig1}, as can be seen by comparing the left and right panels on both figures. The dispersion coefficient $q$ can also be studied as a function of $n'_{d0}$ for both the slow and fast modes, finding that the behaviour of $q$ in this case is very similar to what has been shown in Figs.~\ref{fig1} and~\ref{fig2}.

\begin{figure}[h]\centering
\includegraphics[width=5.8cm]{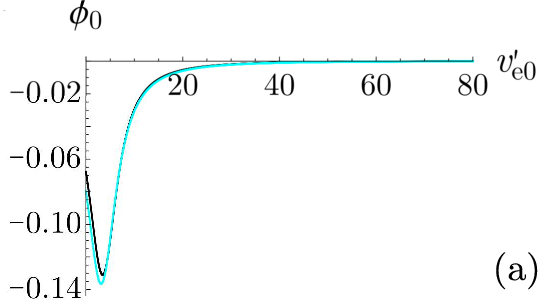}
\includegraphics[width=5.8cm]{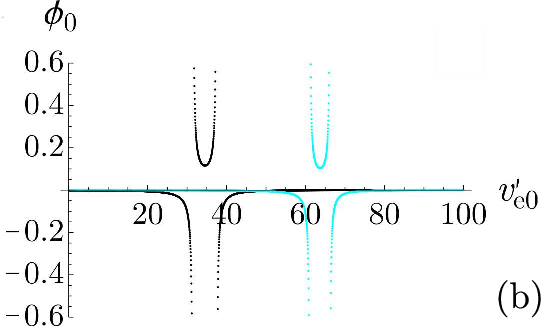}
\includegraphics[width=5.8cm]{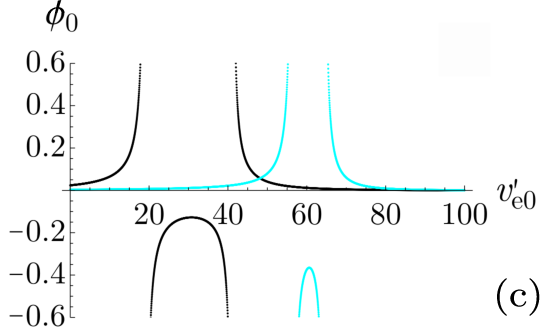}
\caption{(a) Amplitude of soliton $\phi_0$ versus electron initial velocity $v'_{e0}$ for the slow mode, (b) for the fast mode with $Z_d=10$ and (c) for the fast mode  with $Z_d=250$. Black (cyan) dots represent weakly relativistic (relativistic) ions, $v'_{i0}=30$ ($v'_{i0}=60$), $Q_i=0.1$, $n'_{d0}=0.055$, $V=1.0$, and $c'=140$. In (b), divergences are present around $v'_{e0}=32$ and $v'_{e0}=37$ for weakly relativistic ions and around $v'_{e0}=61$ and $v'_{e0}=66$ for relativistic ions. In (c), divergences are present around $v'_{e0}=19$ and $v'_{e0}=42$ for weakly relativistic ions, and around $v'_{e0}=56$ and $v'_{e0}=64$ for relativistic ions.} 
\label{fig5}
\end{figure}

Once the possible values of $q$ have been found for a given set of parameters, we can use those same values to calculate the nonlinear coefficient $p$ in Eq.~\eqref{eq:14}, and thus the soliton amplitude. For the slow mode, results are shown in Fig.~\ref{fig5}(a). In this case, the soliton is always rarefactive, and has a maximum amplitude for normalized electron velocities $v'_{e0}\sim 3$. For the fast mode, as shown in Fig.~\ref{fig1}, there are two branches separated by a forbidden band. For small values of $Z_d$, Fig.~\ref{fig5}(b) is obtained for the soliton amplitude. There is a more complex behavior than for the slow mode [Fig.~\ref{fig5}(a)] and divergences are found, whose positions depend on the ions velocity, as they are given by the condition $B=0$. Also, unlike the slow mode, both compressive and rarefactive solitons exist for the fast mode. From Eqs.~\eqref{eq:14} and~\eqref{eq:17}, it follows that the amplitude can be written as $\phi_0=3V A/B$, whereas the width is $\Delta=\sqrt{4V/A}$. Thus, if $B=0$ the infinite increase of the amplitude occurs while maintaining a finite width. Regarding the fast mode, but for large values of $Z_d$, soliton amplitude is plotted in Fig.~\ref{fig5}(c), where conclusions are analogous to those for small values of $Z_d$ [Fig.~\ref{fig5}(b)].


\section{Relevance of relativistic effects}\label{relativistic}

Here we focus on the effect of relativistic velocities on the soliton properties, by comparing the nonrelativistic, weakly relativistic and fully relativistic cases.

\begin{figure}[h]\centering
\includegraphics[width=6cm]{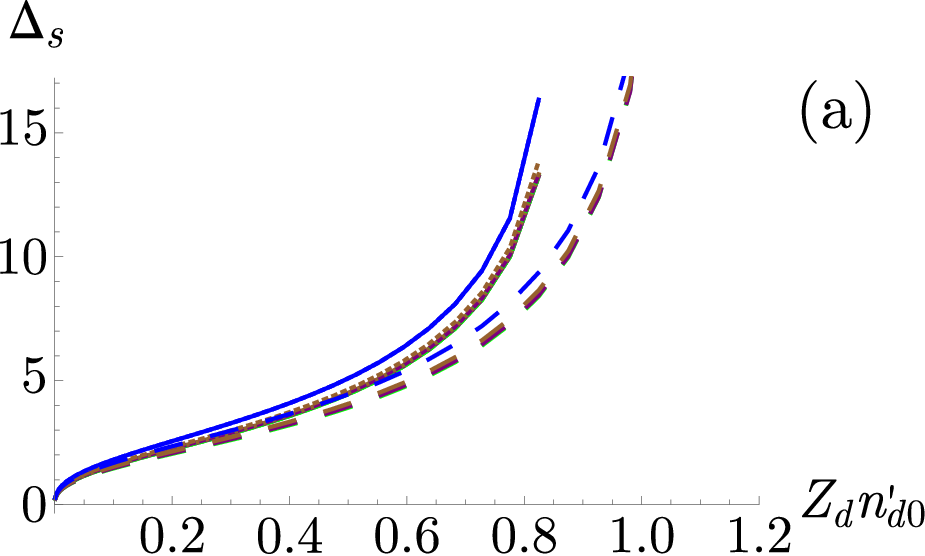}\hspace{0.2cm}
\includegraphics[width=6cm]{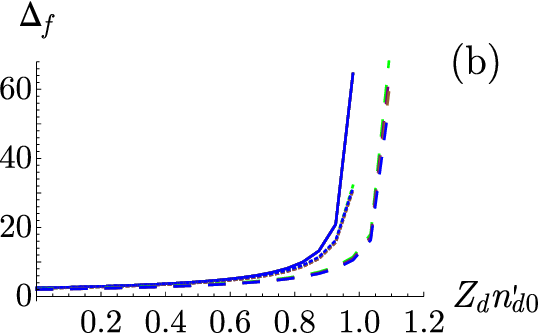}\hspace{0.2cm}
\includegraphics[width=6cm]{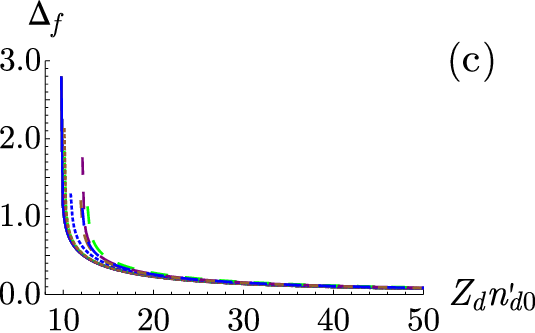}
\caption{Width of the soliton versus dust charge density $Z_d n'_{d0}$ (a) for the slow mode, (b) the fast mode with low charge density, and (c) the fast mode with high charge density. Colors correspond to different electron velocities: $v'_{e0}=10$ (green line), 30 (purple), 50 (brown), 80 (blue). Line types signal the presence of no relativistic effects (solid lines), weakly relativistic effects on ions with $v'_{i0}=30$ (short dashed lines), and strongly relativistic effects on ions with $v'_{i0}=60$ (long dashed lines).}
\label{width_compare}
\end{figure}

We notice from Figs.~\ref{fig1} and \ref{fig2} that the relativistic effects are  more relevant for ions than electrons, thus we will only focus on two values of ion velocity, instead of a range of them. Also, the behaviour of a positive $q$ as a function of $Z_d$ and $n'_{d0}$, is similar, so we will plot the following figures simply as a function of dust charge density $Z_d n'_{d0}$. From Fig. \ref{width_compare} onward, $c'$ is kept at $c'=140$, electron velocities are considered as $v'_{e0}=10$, 30, 50, 80 (green, purple, brown, and blue curves, respectively), and ion velocities as $v'_{i0}=30$, 60. This corresponds to relativistic factors $\beta_0=v'_0/c'$ between $\sim~0.07$ and 0.6 for electrons, and $\sim~0.2$ and~0.4 for ions. In Fig.~\ref{width_compare}, when relativistic effects on ions and electrons are ignored ($\gamma_{i0}=\gamma_{e0}=1$), lines are  continuous; for weakly relativistic effects on ions ($v'_{i0}=30$), short dashed lines are used; for larger relativistic effects ($v'_{i0}=60$), long dashed lines are used. Mass ratio is kept at $Q_i=0.1$, and all other parameters are the same as in Figs.~\ref{fig1}--\ref{fig5}.

First we consider the effect of relativistic velocities on soliton existence and width. We have already shown in Sec.~\ref{KdV} that soliton width is directly related with the dispersion coefficient $q>0$. For the slow mode, Fig.~\ref{width_compare}(a) shows the width as a function of dust charge density. As in Fig.~\ref{fig1} and \ref{fig2}, we only plot the regions where solitons exist. Thus, Fig.~\ref{width_compare}(a) shows that relativistic effects on ions increase the range of dust charge densities where solitons can exist, and that the width decreases when ions become relativistic. Figs.~\ref{width_compare}(b) and \ref{width_compare}(c) are similar to Fig.~\ref{width_compare}(a), but for the fast mode. As seen in Figs~\ref{fig1} and \ref{fig2}, in this case there are two branches, so each plot focuses on one branch, for low and high dust charge density. If the charge density is low [Fig.~\ref{width_compare}(b)], relativistic effects increase the existence range of the soliton, like for the slow mode. 
On the other hand, if the charge density is high [Fig.~\ref{width_compare}(c)], the effects are the opposite. When ions are relativistic, the width increases, and the existence range is decreased.

\begin{figure}[h]\centering
\begin{tabular}{cc}
\includegraphics[width=6.55cm]{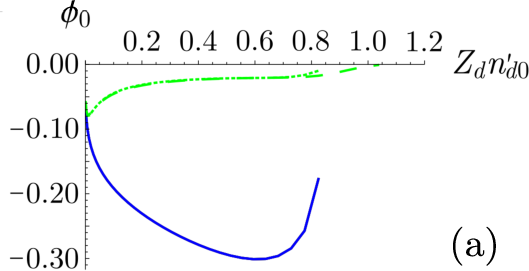} &
\includegraphics[width=6.55cm]{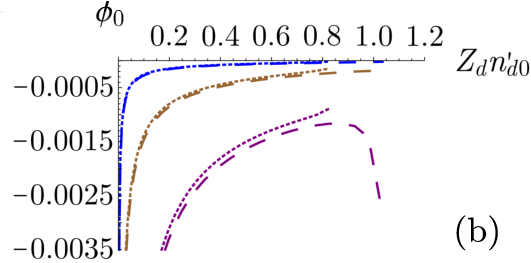} \\
\includegraphics[width=6.55cm]{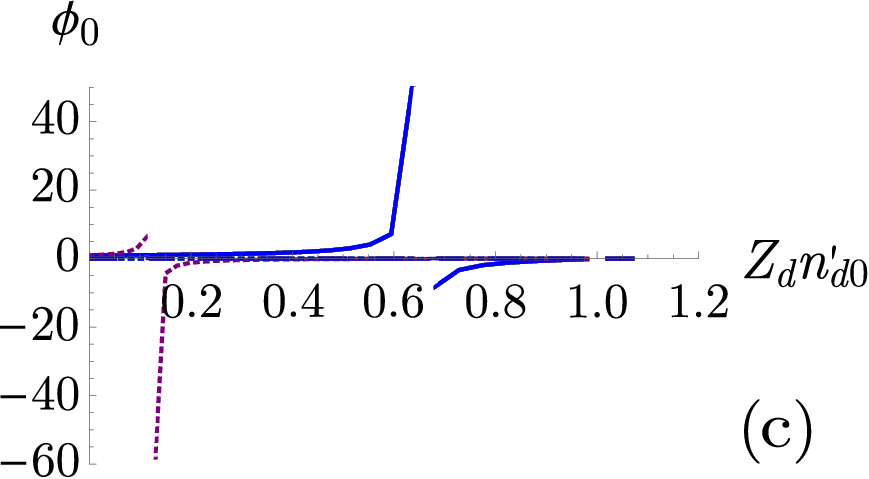} &
\includegraphics[width=6.55cm]{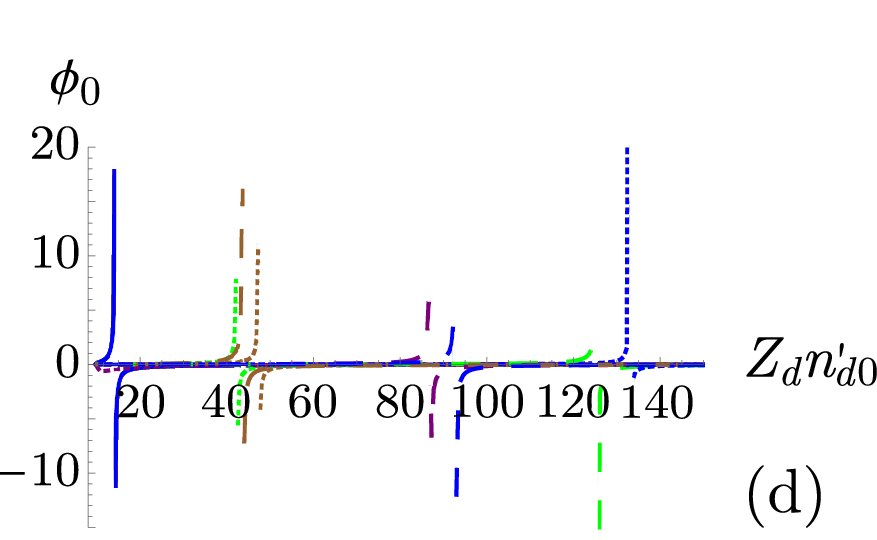}
\end{tabular}
\caption{Same as Fig.~\ref{width_compare}, but (a) and (b) for the amplitude of the slow mode. Due to the large scale difference, the graph has been split in two, as detailed in the text; (c) and (d) for the amplitude of the fast mode with low and high $Z_d n'_{d0}$, respectively.}
\label{amplitude_compare}
\end{figure}

Fig.~\ref{amplitude_compare} shows the effect of relativistic velocities on soliton amplitude, for the slow and fast modes, as a function of dust charge density. For the slow mode, Figs.~\ref{amplitude_compare}(a) and \ref{amplitude_compare}(b), solitons are always rarefactive. In order to better discuss relativistic effects on its amplitude, this figure has been split in two. When relativistic effects are ignored, the amplitude is independent of the electron velocity, and all curves coincide with the blue solid line in Fig.~\ref{amplitude_compare}(a). When relativistic effects are turned on, we have the short dashed lines (weakly relativistic ions) and long dashed lines (strongly relativistic ions). Fig.~\ref{amplitude_compare}(b) shows that this has a strong effect on the soliton amplitude and behavior with dust charge density. In Fig.~\ref{amplitude_compare}(a), for instance, a low electron velocity has been considered ($v_{e0}'=10$, green line). It should be stressed that this change in amplitude is mostly due to the inclusion of relativistic effects in Eqs.~\eqref{eq:1}--\eqref{eq:3}.
Fig.~\ref{amplitude_compare}(b) shows that, once relativistic effects have been established, the amplitude keeps decreasing as electron velocity is increased ($v'_{e0}=30$, purple lines; $v'_{e0}=50$, brown lines; $v'_{e0}=80$, blue lines). Notice the large scale difference between plots~(a) and~(b).

For the fast mode, Figs.~\ref{amplitude_compare}(c) and (d), instead, solitons may be compressive or rarefactive as the dust charge density changes. The amplitude is clearly modified by relativistic effects, mostly due to the location of the discontinuities already observed in Figs.~\ref{fig5}(b) and (c). These discontinuities depend on the location of the zeroes of $B$ in Eqs.~\eqref{eq:15}--\eqref{eq:16}, and it is clear that those locations will change if velocities change, and if relativistic effects are ignored ($\gamma_{j0}=1$).

\begin{figure}[h]\centering
\begin{tabular}{cc}
\includegraphics[width=6.8cm]{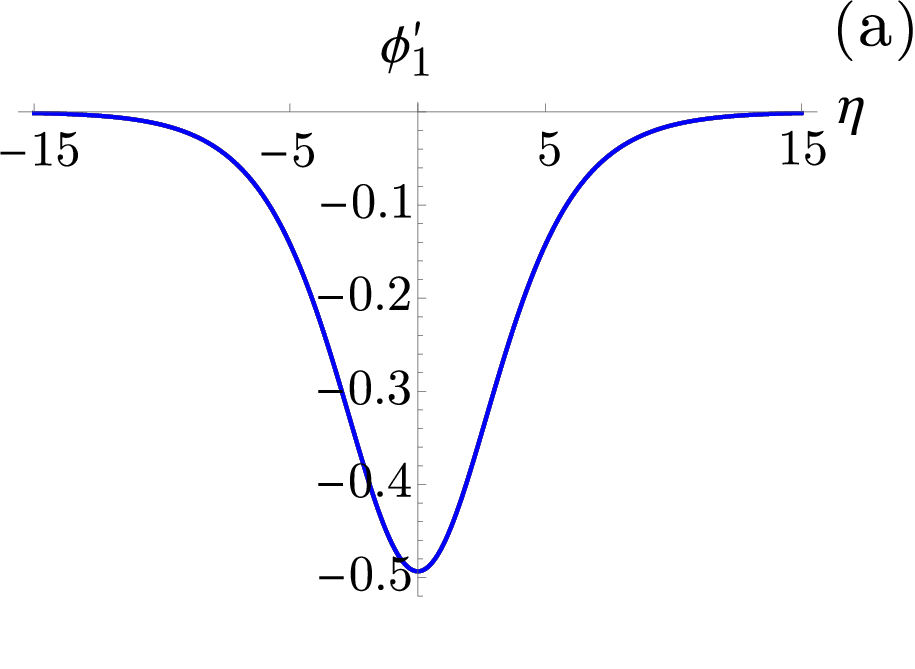} &
\includegraphics[width=6.8cm]{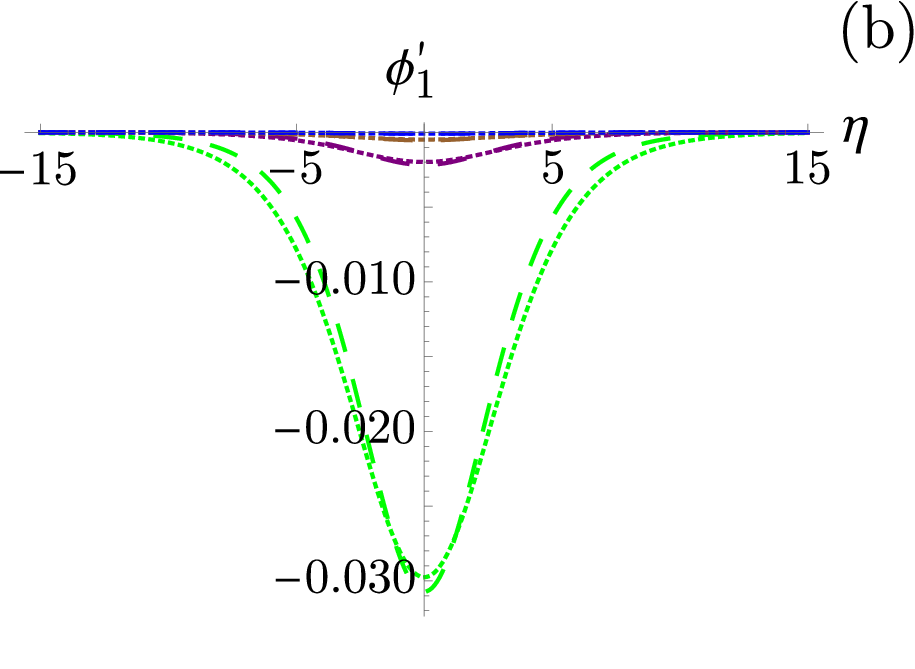}
\end{tabular}
\caption{Soliton profile for the slow mode. Colors correspond to different electron velocities: $v'_{e0}=10$ (green), 30 (purple), 50 (brown), 80 (blue). (a) Without relativistic effects in the model equations~\eqref{eq:1}--\eqref{eq:3}, $v'_{i0}=30$. (b) With relativistic effects, for ion velocity $v'_{i0}=30$ (short dashed line) and $v'_{i0}=60$ (long dashed line).}
\label{profile_s}
\end{figure}

\begin{figure}[h]\centering
\includegraphics[width=5.7cm]{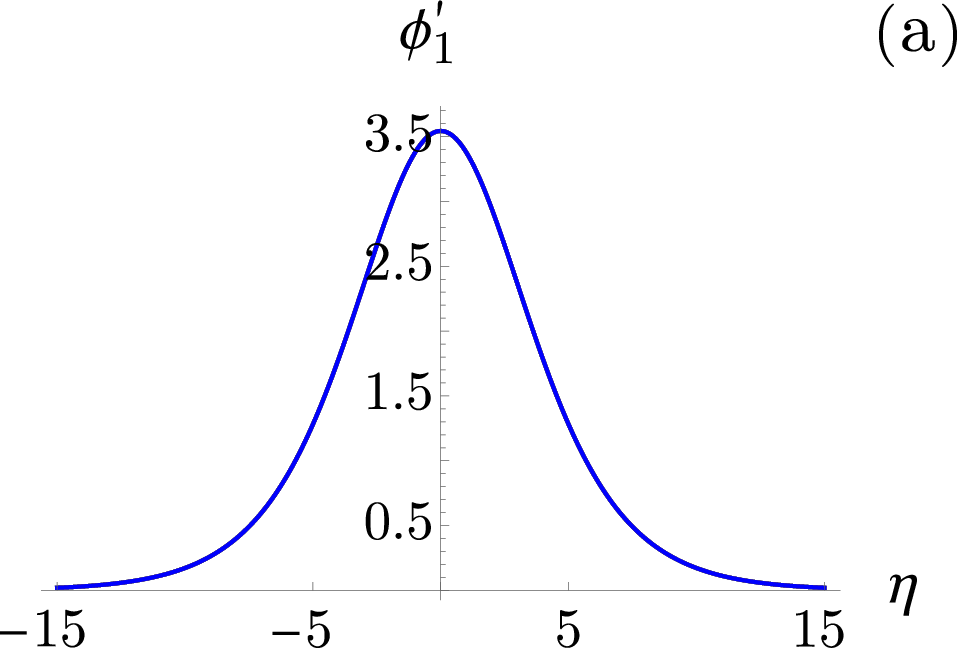} 
\includegraphics[width=5.7cm]{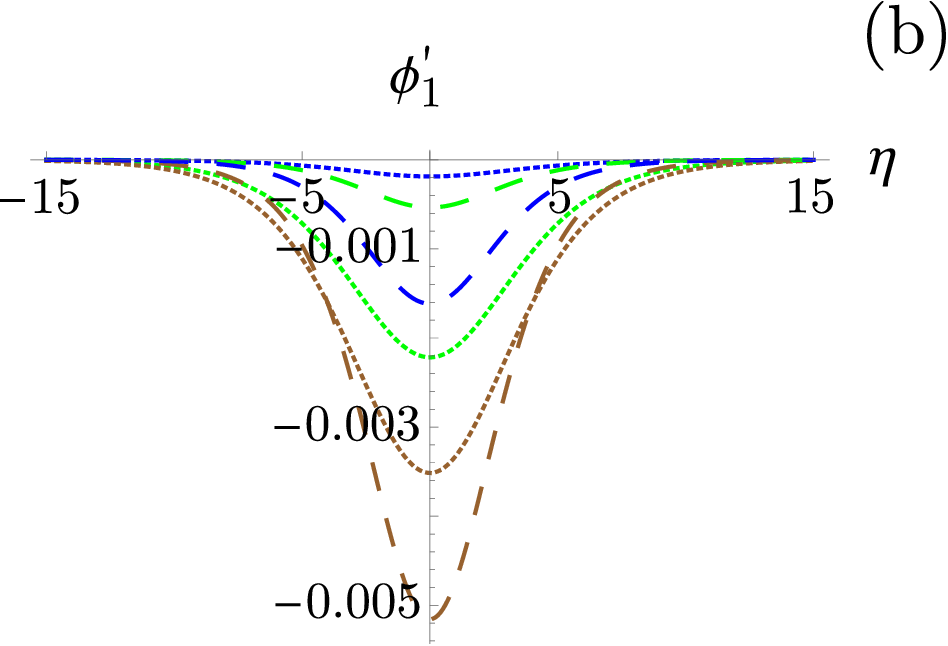}\hspace{0.2cm}
\includegraphics[width=5.7cm]{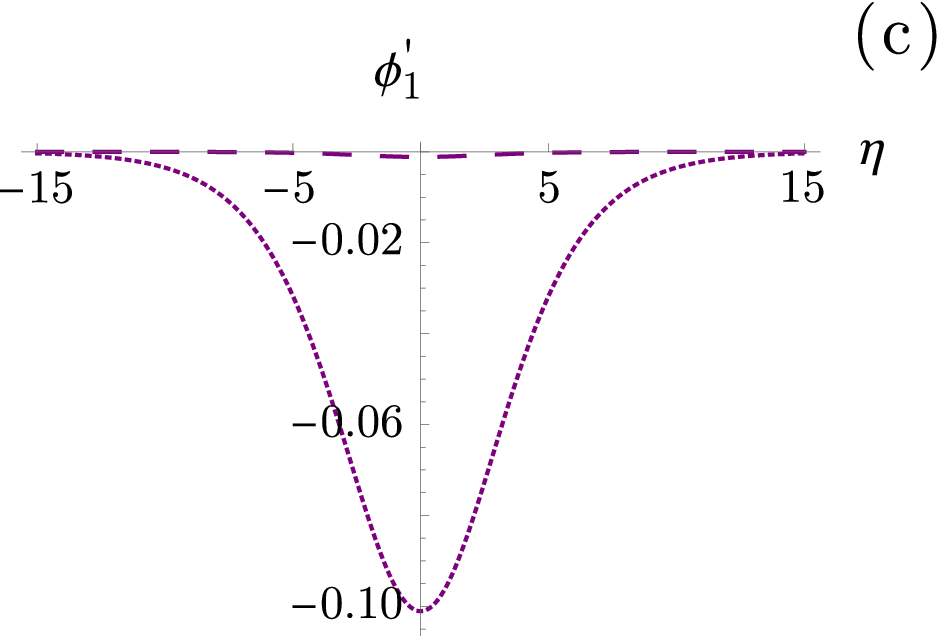}
\caption{Soliton profile for the fast mode and low dust charge density. Here $n'_{d0}=0.055$, $Z_d=10$ and the other parameters are kept the same as in Fig. \ref{profile_s}}
\label{profile_f_small_density}
\end{figure}

\begin{figure}[h]\centering
\begin{tabular}{cc}
\includegraphics[width=8.1cm]{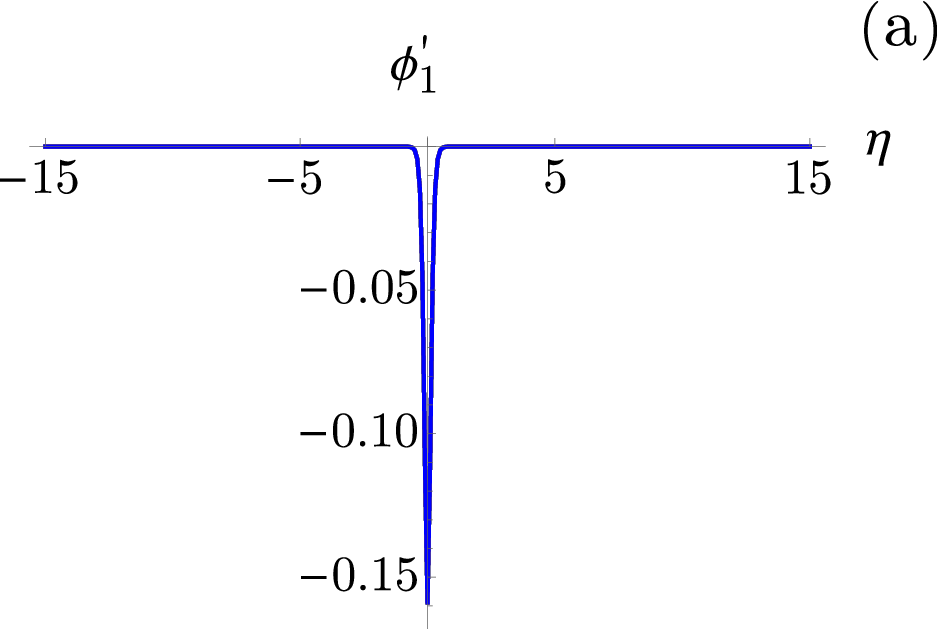}&
\includegraphics[width=8.1cm]{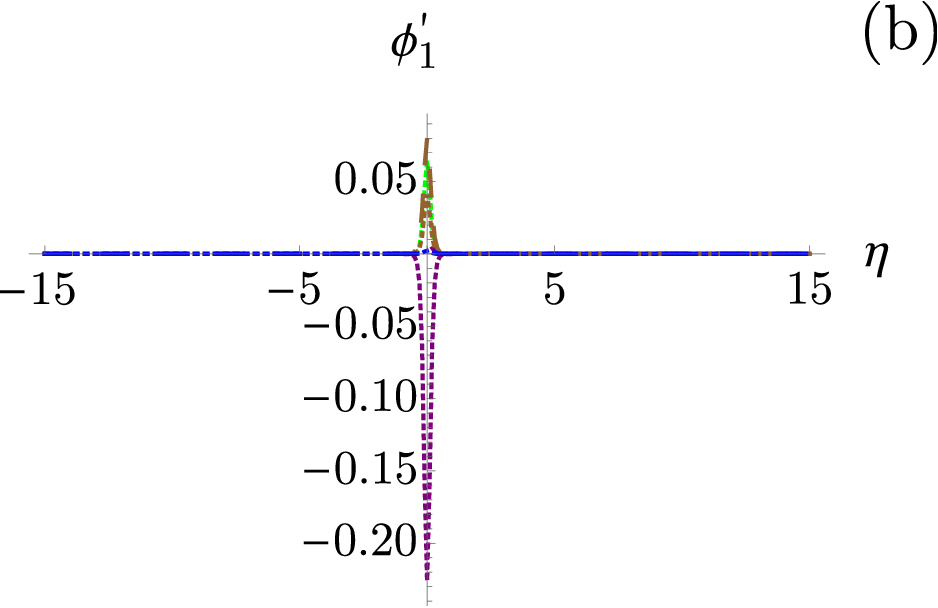}
\end{tabular}
\caption{Soliton profile for the fast mode and high dust charge density. Here $n'_{d0}=0.1$, $Z_d=250$ and the other parameters are kept the same as in Fig. \ref{profile_s}}
\label{profile_f_large_density}
\end{figure}

Soliton profiles for the various modes are shown in Figs.~\ref{profile_s}--\ref{profile_f_large_density}. In Fig.~\ref{profile_s}, for the slow mode, we observe that relativistic effects decrease soliton amplitude, consistent with Fig.~\ref{amplitude_compare}(b). This is a very strong effect, as evidenced by the order of magnitude change between Figs.~\ref{profile_s}(a) and~(b).
However, notice also that, once relativistic effects are taken into account in Eqs.~\eqref{eq:1}--\eqref{eq:3}, increasing ion velocity from $v'_{i0}=30$ to $v'_{i0}=60$ actually increases, slightly, the soliton amplitude. This, of course, depends on the values of velocity that are being considered since the amplitude changes according to that, as can be seen in Fig.~\ref{fig5}. Also, the conditions for soliton amplitude to be positive or negative depending on the velocity of species have been explored in more detail in Ref.~\cite{tesis}.

Regarding the fast mode, the profile in the branch for low dust charge density is shown in Fig.~\ref{profile_f_small_density}. 
It can be observed that, in this case, relativistic effects turn the soliton from compressive to rarefactive. This depends on the exact value of the dust charge density [see Fig.~\ref{amplitude_compare}(c)]. Once relativistic effects are established, the soliton amplitude does not behave monotonically with electron velocity. Notice that for small and large values of the velocity ($v'_{e0}=10$, green and $v'_{e0}=80$, blue), amplitudes are of about the same order, $\phi_1'\sim 0.001$. However, for intermediate values ($v'_{e0}=30$, purple, and $v'_{e0}=50$, brown), amplitude is larger. Actually, in Fig.~\ref{profile_f_small_density}(c), for $v'_{e0}=30$, it is two orders of magnitude larger. Thus, in this parameter region, soliton amplitude is very sensitive to electron velocity, and non-monotonically dependent on it. This complex behavior is consistent with the nontrivial features of soliton amplitude near divergences, as seen in Figs.~\ref{fig5}(b) and \ref{amplitude_compare}(c). The same behavior is observed with respect to ions velocity. In Fig.~\ref{profile_f_small_density}(b) and~(c), it can be seen that changing the ion velocity from weakly relativistic ($v'_{i0}=30$, short dashed lines) to relativistic values ($v'_{i0}=60$, long dashed lines),  the amplitude increases for brown and blue curves, and decreases for green and purple curves.

Finally, we plot the soliton profile for the fast mode and high dust charge density in Fig.~\ref{profile_f_large_density}.
Soliton width is strongly decreased compared to what has been shown for the lower dust charge density branch. Analogous to Fig.~\ref{profile_f_small_density}, solitons may be compressive or rarefactive, depending on the particle velocities, although amplitude changes are more moderate in Fig.~\ref{profile_f_large_density}.


\section{Summary}
\label{summary}

In this work, we have studied solitons in a dusty plasma, considering fully relativistic effects in ion and electron velocities, with nonrelativistic dust particles and temperatures. Relativistic effects have been introduced consistently in the motion equation for the particles. In general, we find that KdV solitons exist, associated to either a fast or a slow phase velocity. The amplitude and width of the soliton, as well as the threshold values, depend on the various physical parameters that we have considered, as can be noticed in Eqs.~\eqref{eq:15}--\eqref{eq:17}. These parameters are the electron and ion velocities, the dust charge density, and ion to dust mass ratio. According to this, the slow mode has a single branch, being present only below a certain threshold, whereas the fast mode has two branches, with a gap where solitons do not exist (Figs.~\ref{fig1} and~\ref{fig2}). It is shown for both modes that relativistic velocities are more relevant for ions than for electrons, and that the effect of the mass ratio can be inferred from Figs.~\ref{fig1} and \ref{fig2} by comparing the left panels (higher ion mass) with the right panels (lower ion mass), observing that it modifies the limits for soliton existence.

It can be seen that soliton width increases with dust charge density for the slow mode [Fig.~\ref{width_compare}(a)]. As to the fast mode, the two branches have different behaviors: before the gap, behavior is similar to the slow mode [Fig.~\ref{width_compare}(b)], but after the gap, it is the opposite: soliton width decreases with dust charge density [Fig.~\ref{width_compare}(c)].

Soliton amplitude and profiles have also been studied for the slow mode [Figs.~\ref{fig5}(a), ~\ref{amplitude_compare}(a), ~\ref{amplitude_compare}(b) and~\ref{profile_s}], and for the two branches of the fast mode [Figs.~\ref{fig5}(b),~\ref{fig5}(c),~\ref{amplitude_compare}(c),~\ref{amplitude_compare}(d),~\ref{profile_f_small_density} and~\ref{profile_f_large_density}].
For the slow mode, relativistic velocities on the ions have only a minor effect, and the soliton is always rarefactive, having a maximum amplitude, and then monotonically decreasing its amplitude for larger values of $v'_{e0}$. 
The fast mode exhibits a more complex behavior, since compressive and rarefactive solitons can be found. This depends on dust charge density, and electron and ion velocity. There are divergences in the amplitude at certain values, which mark the transition from compressive to rarefactive and vice versa. As ion velocity increases to relativistic values, the divergence appears at larger values of $v'_{e0}$. 

It should be taken into account, though, that plasmas can be subject to various instabilities, which could affect the actual existence of solitons in the system. This should need a separate analysis of the parameter space, {\it e.g.\/} by means of numerical simulations. Also, in this paper we have considered fully relativistic effects on the particle beam velocities. However, thermal velocities can approach the velocity of light for large temperatures too, thus future work would need to consider the effect of relativistic temperatures. This modifies the thermal term in the motion equation \eqref{eq:2}, and needs the introduction of the enthalpy to effectively  replace the mass (see, {\it e.g.\/} Ref.~\cite{Asenjo_a}).
Considering the case of an adiabatic plasma should be relevant as well. This would lead to include a further Lorentz factor in the pressure term in Eq.~\eqref{eq:2}, which could modify existence range and amplitude of the solitons, as well as the location of divergences observed in Figs.~\ref{fig5} and~\ref{amplitude_compare}.

Future work should also take into account the presence of a magnetic field, since it is especially relevant for astrophysical systems. This would modify the motion equations and break the spatial symmetry; thus the soliton properties are expected to depend on the propagation angle with respect to the magnetic field. 


\begin{acknowledgements}
 This project has been financially supported by FONDECyT under contracts No. 1161711 and 1201967 (V.M.) and by CONICyT through a Doctoral Fellowship, Contract No. 21161594 (M.C.).  
\end{acknowledgements}

\bibliography{soliton_dust_FAZ_Aug18}
\bibliographystyle{h-physrev}

\end{document}